\documentclass[nofootinbib,aps,a4paper,letterpaper,superscriptaddress,
twocolumn, times,eqsecnum]{revtex4}
\usepackage{amsmath}
\usepackage{amsfonts}
\usepackage{booktabs}
\usepackage{multirow}
\usepackage{siunitx}
\usepackage{adjustbox}
 \pdfoutput=1
\usepackage{graphicx}
\usepackage{color}
\usepackage{braket}
\usepackage{dcolumn}
\usepackage{bm,url}
\usepackage[linktocpage]{hyperref}
\usepackage{subfigure}
\usepackage{amsfonts}
\usepackage{orcidlink}
\usepackage[usenames,dvipsnames,svgnames]{xcolor}  
\usepackage{hyperref}
\definecolor{oxfordblue}{rgb}{0.0, 0.13, 0.28}
\definecolor{burgundy}{rgb}{0.5, 0.0, 0.13}
\definecolor{darkolivegreen}{rgb}{0.33, 0.42, 0.18}
\definecolor{darkblue}{rgb}{0,0,0.5}
\definecolor{richcarmine}{rgb}{0.84, 0.0, 0.25}
\definecolor{darkblue}{rgb}{0,0,0.5}
\definecolor{bluer}{rgb}{0.00,0.50,0.75}{}
\hypersetup{colorlinks=true, citecolor=red, linkcolor=blue,
 urlcolor = magenta, filecolor=magenta}

\begin{document}

\title{Reconstructing the generalized Barrow holographic dark energy with
physics-informed neural networks}

\author{Spyros Basilakos\
,\orcidlink{0000-0001-5066-0259}
}
\email{svasil@academyofathens.gr}
\affiliation{National Observatory of Athens, Lofos Nymfon, 11852 Athens, Greece}
\affiliation{Academy of Athens, Research Center for Astronomy and Applied
Mathematics, Soranou Efesiou 4, 11527, Athens, Greece}
\affiliation{School of Sciences, European University Cyprus, Diogenes Street,
Engomi, 1516 Nicosia, Cyprus}

\author{Andronikos Paliathanasis
,\orcidlink{0000-0002-9966-5517}
}
\email{anpaliat@phys.uoa.gr}
\affiliation{Institute of Systems Science, Durban University of Technology,
Durban 4000, South Africa}
\affiliation{Centre for Space Research, North-West University, Potchefstroom
2520, South Africa. }
\affiliation{Centro de Investigaci\'on, Innovaci\'on y Creaci\'on (CIIC),
Universidad
Cat\'olica de Temuco, Temuco, Chile}
\affiliation{Departamento de Ciencias Matem\'{a}ticas y F\'{\i}sicas, Facultad
de Ingenier\'{\i}a, Universidad Cat\'olica de Temuco, Temuco, Chile}
\affiliation{National Institute for Theoretical and Computational Sciences
(NITheCS),
South Africa}

\author{Emmanuel N. Saridakis
,\orcidlink{0000-0003-1500-0874}
}
\email{msaridak@noa.gr}
 \affiliation{Institute for Astronomy, Astrophysics, Space Applications and
Remote Sensing, National Observatory of Athens, 15236 Penteli, Greece}
 \affiliation{Departamento de Matem\'{a}ticas, Universidad Cat\'{o}lica del
  Norte, Avda. Angamos 0610, Casilla 1280, Antofagasta, Chile}
 \affiliation{CAS Key Laboratory for Research in Galaxies and Cosmology,
School
  of Astronomy and Space Science,
  University of Science and Technology of China, Hefei 230026, China}

\author{Stylianos A. Tsilioukas
,\orcidlink{0009-0003-3051-3405}
}
\email{tsilioukas@sch.gr}
\affiliation{Institute for Astronomy, Astrophysics, Space Applications and
Remote Sensing, National Observatory of Athens, 15236 Penteli, Greece}

\begin{abstract}
Barrow holographic dark energy connects cosmic acceleration with possible
quantum-gravitational deformations of horizon entropy, encoded in the Barrow
exponent $\Delta$. If such effects are scale dependent, however, there is no
fundamental reason for $\Delta$ to remain constant throughout cosmic history.
In this work we reconstruct $\Delta(z)$ directly from observations, without
assuming any particular functional form, using the Cosmo-PINN
physics-informed neural-network framework. The generalized Barrow
holographic evolution equation is incorporated into the training, while
PantheonPlus supernovae, DESI DR2 baryon acoustic oscillations and cosmic
chronometers constrain the reconstruction. We find a mild and smooth
redshift evolution, with the posterior mean favoring negative $\Delta$ and
this tendency becoming stronger when the Cepheid calibration is included.
Nevertheless, $\Delta=0$ and constant negative values remain compatible with
current uncertainties. The reconstructed cosmology yields a viable late-time
evolution, with $w_{\rm DE}$ close to $-1$ and the expected transition to
accelerated expansion. Our results demonstrate that cosmological observations
can directly probe the functional behavior of a quantity entering the
underlying entropy law itself.
\end{abstract}

\maketitle

\section{Introduction}

The late-time accelerated expansion of the Universe remains one of the central
problems of modern cosmology. Although the cosmological constant within the
$\Lambda$CDM paradigm provides the simplest and phenomenologically successful
description, its extremely small observed value compared to theoretical
expectations, together with the possibility of a dynamical origin of cosmic
acceleration, has motivated a wide range of alternatives. These include both
dynamical dark-energy scenarios within general relativity
\cite{Copeland:2006wr,
Cai:2009zp, Bamba:2012cp} and modifications of the gravitational sector
\cite{Nojiri:2010wj,Capozziello:2011et,CANTATA:2021ktz}.
Moreover, recent high-precision observations, particularly the DESI baryon
acoustic oscillation measurements, have strengthened the motivation for
investigating dynamical dark-energy behavior beyond a strictly constant
vacuum-energy contribution \cite{DESI:2025zgx}.

A conceptually different possibility is holographic dark energy, which is based
on the holographic principle and the connection between gravity, horizon
thermodynamics and quantum field theory. In its standard formulation, requiring
that the energy contained in a region of characteristic size $L$ does not
exceed
the mass of a black hole of the same size leads to a dark-energy density
controlled by an infrared cutoff \cite{Li:2004rb,Wang:2016och}. The resulting
sector is intrinsically dynamical and directly connects cosmological evolution
with horizon thermodynamics. This holographic dark energy has been shown to
lead to
rich and interesting phenomenology
\cite{Li:2004rb,Wang:2016och,Huang:2004ai,Pavon:2005yx,
Wang:2005jx,Nojiri:2005pu,Kim:2005at,Setare:2006wh,Sheykhi:2009dz,
Li:2009bn,
Zhang:2009un,Micheletti:2010cm,Duran:2010hi,Micheletti:2009jy,
Aviles:2011sfa,Zhai:2011pp,
Cardenas:2010wx,Chimento:2011pk,Zhang:2012uu,Cruz:2011wx,Huang:2012xma,
Zhang:2015rha,Nastase:2016sji,Mukherjee:2016lor,Zhao:2017urm,
Luongo:2017yta,Mukherjee:2017oom,
Pinki:2022aht,Mohammadi:2022vru,Landim:2022jgr,Trivedi:2024inb,Maity:2024tkq,
Li:2024bwr}. Such   phenomenological success led to the investigation of its
extensions, either by changing the  infrared cutoff
\cite{Gong:2004fq,Cai:2007us,Setare:2008bb,Gong:2009dc,
Suwa:2009gm,Bouhmadi-Lopez:2011qvd,
Malekjani:2012bw,Landim:2015hqa,Rao:2017lnt,Shekh:2021ule,Rudra:2022qbv}, or by
considering modified entropy-area
relations
\cite{Pasqua:2015bfz,Jawad:2016tne,Pourhassan:2017cba,Saridakis:2017rdo,
Nojiri:2017opc,Saridakis:2018unr,DAgostino:2019wko,
Dabrowski:2020atl,daSilva:2020bdc,
Bhattacharjee:2020ixg, Drepanou:2021jiv,Hossienkhani:2021emv,
Nojiri:2021iko,Jusufi:2021fek,Hernandez-Almada:2021aiw,
Luciano:2022pzg,Yarahmadi:2024oqv,Cimdiker:2025vfn,Luciano:2026vhm,
Luciano:2025ovj,Halder:2026wvg}.

 One particularly interesting case is Barrow entropy \cite{Barrow:2020tzx},
which incorporates possible quantum-gravitational deformations.
It is given by $S\propto A^{1+\Delta/2}$, with $A$ the horizon area and where
 the exponent $\Delta$ quantifies the horizon deformation and thus the
departure from standard Bekenstein-Hawking entropy, which is recovered for
  $\Delta=0$. Let us mention that the possibility that quantum-gravitational
effects can drive the late-time
acceleration has also been explored through different routes, such as the
topological dark energy at the spacetime-foam
level \cite{Tsilioukas:2023tdw,
Anagnostopoulos:2025tax} or through Wald-Gauss-Bonnet
entropy \cite{Tsilioukas:2024seh,Petronikolou:2025mlm,Tsilioukas:2026gvy}.  The
implementation of Barrow entropy in the holographic framework gives rise to
Barrow holographic dark energy \cite{Saridakis:2020zol}, characterized by
$\rho_{DE}\propto L^{\Delta-2}$ and exhibiting a rich cosmological
behavior \cite{Saridakis:2020zol,Anagnostopoulos:2020ctz,Mamon:2020spa,
Srivastava:2020cyk,Chakraborty:2021uzp,Adhikary:2021xym,Rani:2021hvh,
Huang:2021zgj,Nandhida:2021vxl, Nojiri:2021jxf,Maity:2022gdy,
Paul:2022doh,Kumar:2022acs,Saleem:2022eti,Remya:2022frs,Luciano:2022viz,
 Luciano:2022ffn, Luciano:2022hhy,Boulkaboul:2023yks,
Mahmoudifard:2024gmn,Luciano:2025ykr,Yarahmadi:2024afr,Luciano:2025elo,
Maity:2025cxd, Yarahmadi:2025ujq}.

There is, however, no fundamental reason for the Barrow exponent to remain
constant throughout cosmic history, since energy-scale dependence is common
in quantum field theory and quantum-gravity settings. Hence, since $\Delta$
parametrizes quantum-gravitational modifications of the horizon, their
magnitude
may depend on the physical scale. In an expanding Universe this translates into
an effective time, or redshift, dependence. This possibility motivated Barrow
holographic dark energy with varying exponent \cite{Basilakos:2023seo}, where
several phenomenological forms for $\Delta(z)$ were considered and shown to
generate viable and richer cosmological behavior.

Once the exponent $\Delta$ is allowed to evolve, the question is
  what determines its evolution form. Prescribing linear, CPL-like,
exponential, hyperbolic-tangent, or other ans\"atze is useful for exploring the
phenomenology \cite{Basilakos:2023seo}, but inevitably imprints the assumed
functional form on the results. Hence, the natural next step is to reverse the
problem and study whether $\Delta(z)$ itself can be reconstructed directly from
cosmological observations, while requiring it to satisfy the equations of
Barrow
holographic dark energy.

Data-driven reconstruction techniques offer a promising route toward this goal
\cite{Yang:2015tzc,Seikel:2012uu,Mu:2023bsf,Dialektopoulos:2023dhb,
Mitra:2024ahj,Alda:2026nlb}. Gaussian Processes and artificial neural networks can
reconstruct
cosmological functions without imposing a low-dimensional parametrization, but
a
purely data-driven reconstruction does not guarantee consistency with the
dynamical equations of the underlying theory. This is crucial here, since
$\Delta(z)$ is not an arbitrary phenomenological function, but rather it enters
the holographic energy density and the cosmological evolution equations,
together with its derivative. A reconstruction that ignores these relations
could therefore fit the observations without corresponding to a physically
consistent Barrow holographic cosmology.

Physics-Informed Neural Networks (PINNs) provide a natural way to overcome this
limitation \cite{Raissi:2017zsi}. By incorporating the governing differential
equations directly into the learning process, they constrain the reconstructed
solution simultaneously by observations and by the physical laws of the system.
This approach has recently been developed for cosmological reconstruction
through the Cosmo-PINN framework \cite{Paliathanasis:2026dqk}, in which the
cosmological equations are embedded into the loss function, ensuring dynamical
consistency throughout the training domain. For various applications of PINNs
in cosmology we refer the reader to
\cite{Dialektopoulos:2026jmw,Chantada:2022bdf,Yarahmadi:2025luc,Verma:2025ujt}
and references therein.

In the present work we apply this strategy to Barrow holographic dark energy
with varying exponent. Instead of imposing a particular functional form for
$\Delta(z)$, we promote the Barrow exponent itself to a quantity reconstructed
by the Cosmo-PINN. The observational data determine its redshift evolution,
while the Barrow holographic dark-energy equations constrain the training to
physically admissible cosmological histories. The framework therefore combines
the flexibility of a non-parametric reconstruction with the theoretical
information encoded in the holographic model, allowing us to examine whether
observations favor a constant deformation, a genuinely evolving exponent, or an
evolution toward the standard entropy limit without selecting any of these
possibilities beforehand.

The novelty of the analysis is twofold. From the Barrow holographic dark-energy
perspective, we replace the previously prescribed evolution of the
quantum-gravitational exponent by a data-driven reconstruction. From the
machine-learning perspective, the physics-informed framework is used not simply
to reconstruct a standard cosmological observable such as $H(z)$ or
$w_{DE}(z)$,
but to infer a function characterizing the underlying horizon entropy. The
analysis thus establishes a direct link between late-time cosmological
observations and the effective evolution of the entropy deformation, allowing
the data to probe the functional behavior of a quantity associated with the
quantum-gravitational structure of the model.

The plan of the work is the following. In Section~\ref{model} we review
Barrow holographic dark energy and construct the cosmological equations for a
varying Barrow exponent. In Section~\ref{sec3} we present the Cosmo-PINN
implementation for reconstructing $\Delta(z)$. In Section~\ref{sec4} we
present the reconstruction and its cosmological implications, while
Section~\ref{Conclusions} is devoted to the conclusions.

\section{Generalized Barrow Holographic Dark Energy}
\label{model}

In this section we present the cosmological framework of Barrow holographic
dark
energy with a varying exponent, which will constitute the theoretical basis of
the reconstruction performed in the following sections. We start from Barrow
entropy \cite{Barrow:2020tzx},
\begin{equation}
\label{eq:Barrow_entropy}
S_{B}=\left(\frac{A}{A_{0}}\right)^{1+\Delta/2},
\end{equation}
where $A$ denotes the horizon area, $A_{0}$ is the Planck area, and $\Delta$ is
the Barrow exponent that measures the deviation from the standard
Bekenstein-Hawking entropy. In particular, the usual area law is recovered for
$\Delta=0$.

{{
In Barrow's original geometrical construction, the exponent is
restricted to $0\leq\Delta\leq1$. However, as discussed in
\cite{Jizba:2024klq,Luciano:2025hjn},
this restriction is associated with the particular fractal
geometry considered in the original proposal and can be
relaxed in more general settings. In particular, porous
fractal geometries with effective dimensions smaller than
two, as well as negative anomalous scaling dimensions,
provide motivation for considering $\Delta<0$.
In the present work, aiming at a general reconstruction,
we do not impose the original geometrical restriction,
but allow both positive and negative values of $\Delta$.
Negative values should therefore be interpreted within this broader phenomenological
extension of Barrow entropy, rather than within Barrow's original geometrical
construction.}}

When the entropy relation (\ref{eq:Barrow_entropy}) is incorporated into the
holographic dark-energy construction, the corresponding energy density acquires
the form \cite{Saridakis:2020zol}
\begin{equation}
\label{eq:BHDE_density}
\rho_{DE}=C L^{\Delta-2},
\end{equation}
where $L$ is the infrared cutoff and $C$ is the model parameter. For
$\Delta=0$,
Eq.~(\ref{eq:BHDE_density}) reduces to the standard holographic expression
$\rho_{DE}=3c^{2}M_{p}^{2}L^{-2}$, with $C=3c^{2}M_{p}^{2}$.

We consider a spatially flat Friedmann-Robertson-Walker geometry,
\begin{equation}
\label{eq:FRW_metric}
ds^{2}=-dt^{2}+a^{2}(t)\delta_{ij}dx^{i}dx^{j},
\end{equation}
and choose the future event horizon as the infrared cutoff, namely
\cite{Li:2004rb}
\begin{equation}
\label{eq:event_horizon}
L=R_{h}\equiv a(t)\int_{t}^{\infty}\frac{dt'}{a(t')}
=a\int_{a}^{\infty}\frac{da'}{H(a')a'^{2}},
\end{equation}
where $H\equiv\dot a/a$. Hence, the Barrow holographic dark-energy density
becomes
\begin{equation}
\label{eq:rhoDE_Rh}
\rho_{DE}=C R_{h}^{\Delta-2}.
\end{equation}
The   Friedmann equations are then
\begin{eqnarray}
\label{eq:Friedmann1}
3M_{p}^{2}H^{2}&=&\rho_{m}+\rho_{DE},\\
\label{eq:Friedmann2}
-2M_{p}^{2}\dot H&=&\rho_{m}+p_{m}+\rho_{DE}+p_{DE},
\end{eqnarray}
with $p_{DE}$ the pressure of holographic dark energy.
In the following we restrict ourselves to pressureless matter, $p_m=0$, which
is
separately conserved and therefore satisfies $\rho_m=\rho_{m0}a^{-3}$.
Introducing the usual density parameters
\begin{equation}
\label{eq:density_parameters}
\Omega_m\equiv\frac{\rho_m}{3M_p^2H^2},
\qquad
\Omega_{DE}\equiv\frac{\rho_{DE}}{3M_p^2H^2},
\end{equation}
the first Friedmann equation implies $\Omega_m+\Omega_{DE}=1$. Consequently,
\begin{equation}
\label{eq:Hubble_Omega}
\frac{H^{2}}{H_{0}^{2}}
=
\frac{\Omega_{m0}a^{-3}}{1-\Omega_{DE}},
\end{equation}
or, equivalently, in terms of the redshift,
\begin{equation}
\label{eq:Hubble_redshift}
\left[\frac{H(z)}{H_{0}}\right]^{2}
=
\frac{\Omega_{m0}(1+z)^{3}}{1-\Omega_{DE}(z)}.
\end{equation}

{{
We now generalize the construction by allowing the Barrow
exponent to evolve throughout the cosmological history,
namely $\Delta=\Delta(z)$, following the theoretical
framework developed in Ref.~\cite{Basilakos:2023seo}. In this
framework, the standard Friedmann equations are retained,
with the future event horizon serving as the infrared
cutoff of the effective holographic dark-energy component.
This differs from the thermodynamic approach of
Ref.~\cite{Nojiri:2021jxf}, where modified Friedmann equations
are derived by applying the first law of thermodynamics
to the varying Barrow entropy at the apparent horizon.
In the present framework, the variation of $\Delta$ is
consistently incorporated into the cosmological evolution
equations, as explicitly derived in Ref.~\cite{Basilakos:2023seo}.
In particular, the resulting equations contain terms
proportional to the derivative of the Barrow exponent.
}}

Using $ \ln a=-\ln(1+z)$ and the relation
$
\frac{d}{d \ln a}=-(1+z)\frac{d}{dz}$,
the evolution equation can be expressed directly in redshift space as
\begin{eqnarray}
\label{eq:OmegaDE_evolution}
&&
\!\!\!\!\!\!\!\!\!
\frac{(1+z)}{\Omega_{DE}(1\!-\!\Omega_{DE})}
\frac{d\Omega_{DE}}{dz}
=
\frac{1+z}{\Delta-2}\frac{d\Delta}{dz}
\ln\left[
\frac{1-\Omega_{DE}}{\Omega_{DE}}P(z)
\right]\nonumber\\
&&  \!\!\!\!\!\!\!\!\!
-(2-\Delta)
\sqrt{\frac{\Omega_{DE}}{C}}
\left[
\frac{1\!-\!\Omega_{DE}}{\Omega_{DE}}P(z)
\right]^{\frac{\Delta}{2(\Delta-2)}}
+\Delta+1,
\end{eqnarray}
where
\begin{equation}
\label{eq:Pz}
P(z)\equiv
\frac{C}{H_{0}^{2}\Omega_{m0}}(1+z)^{3}.
\end{equation}
Here and in what follows we employ the normalization used in the numerical
reconstruction, in which the corresponding powers of $3M_p^2$ are absorbed into
the definition of $C$. Equation~(\ref{eq:OmegaDE_evolution}) is the central
background equation of the generalized scenario: once $\Delta(z)$ is specified,
or reconstructed, it determines the evolution of $\Omega_{DE}(z)$ and, through
Eq.~(\ref{eq:Hubble_redshift}), the expansion history.

For completeness, we also derive the effective equation-of-state parameter of
the dark-energy sector. Differentiating Eq.~(\ref{eq:rhoDE_Rh}) for a varying
$\Delta$ gives
\begin{equation}
\label{eq:rhodot}
\dot{\rho}_{DE}
=
\rho_{DE}
\left[
\dot{\Delta}\ln R_h
+
(\Delta-2)\frac{\dot R_h}{R_h}
\right],
\end{equation}
while from Eq.~(\ref{eq:event_horizon}) one has
\begin{equation}
\label{eq:Rhdot}
\dot R_h=H R_h-1.
\end{equation}
Here and in the following, the horizon radius is understood
to be expressed in units of a fixed reference length $L_*$,
such that the logarithm $\ln R_h$ denotes
$\ln(R_h/L_*)$. This convention is also adopted in the
numerical reconstruction.
Therefore, using the dark-energy conservation equation
\begin{equation}
\label{eq:DE_conservation}
\dot{\rho}_{DE}+3H\rho_{DE}(1+w_{DE})=0,
\end{equation}
we obtain
{\small{
\begin{eqnarray}
\label{eq:wDE_redshift}
w_{DE}(z)
&=&
-\frac{1+\Delta}{3}
+\frac{\Delta-2}{3}
\sqrt{\frac{\Omega_{DE}}{C}}
\left[
\frac{1\!-\!\Omega_{DE}}{\Omega_{DE}}P(z)
\right]^{\frac{\Delta}{2(\Delta\!-\!2)}}
\nonumber\\
&&
+\frac{1+z}{3(\Delta-2)}
\frac{d\Delta}{dz}
\ln\left[
\frac{1\!-\!\Omega_{DE}}{\Omega_{DE}}P(z)
\right].
\end{eqnarray}}}
Thus, both the value of the Barrow exponent and its redshift variation affect
the effective dark-energy equation of state. In the constant-exponent limit,
$d\Delta/dz=0$, the standard Barrow holographic dark-energy equations are
recovered, while for $\Delta=0$ one further obtains ordinary holographic dark
energy \cite{Li:2004rb,Saridakis:2020zol,Wang:2016och}.

Equations~(\ref{eq:Hubble_redshift}), (\ref{eq:OmegaDE_evolution}), and
(\ref{eq:wDE_redshift}) form the background system that will be implemented in
the physics-informed reconstruction. Contrary to previous analyses, where a
particular functional dependence for $\Delta(z)$ was imposed beforehand, in the
present work no phenomenological ansatz for its redshift evolution is assumed.
Instead, $\Delta(z)$ is treated as the function to be inferred from the
observational data, while Eq.~(\ref{eq:OmegaDE_evolution}) is imposed as the
physical constraint of the Cosmo-PINN framework described in the next section.

\section{Cosmo-PINN reconstruction framework}
\label{sec3}

We now present the Cosmo-PINN framework \cite{Paliathanasis:2026dqk}, which we
employ
to reconstruct the generalized Barrow holographic dark-energy scenario
directly from cosmological observations. Cosmo-PINN is a
Physics-Informed Neural Network (PINN) in which the cosmological
dynamical equations are incorporated into the training procedure as
physical constraints. Hence, in contrast to a purely data-driven neural
network, the reconstructed functions are required not only to reproduce
the observational data, but also to satisfy the underlying cosmological
dynamics. In the present case, the quantity of primary interest is the
Barrow exponent $\Delta(z)$, while the evolution equation
(\ref{eq:OmegaDE_evolution}) is imposed throughout the training domain.

For the reconstruction we use the following late-time background
observational datasets:
\begin{itemize}

\item \textit{Type Ia supernovae (SNIa):} We employ the PantheonPlus
compilation \cite{Brout:2022vxf}, which contains 1550 Type Ia supernova
measurements of the distance modulus $\mu^{\rm obs}$ over the redshift
range $10^{-3}<z<2.27$. We perform the analysis both without the
Cepheid calibration, hereafter denoted as PP, and including the Cepheid
calibration, denoted as PPS.

\item \textit{Baryon acoustic oscillations (BAO):} We use the recent
measurements from the Dark Energy Spectroscopic Instrument Data Release 2
(DESI DR2) \cite{DESI:2025zpo,DESI:2025zgx,DESI:2025fii}. The dataset
provides measurements of the transverse comoving-distance ratio
\begin{equation}
\frac{D_M}{r_{\rm drag}}
=
\frac{D_L}{(1+z)r_{\rm drag}},
\end{equation}
the volume-averaged distance ratio
\begin{equation}
\frac{D_V}{r_{\rm drag}}
=
\frac{\left(zD_HD_M^2\right)^{1/3}}{r_{\rm drag}},
\end{equation}
and the Hubble-distance ratio
\begin{equation}
\frac{D_H}{r_{\rm drag}}
=
\frac{c}{r_{\rm drag}H(z)},
\end{equation}
at seven effective redshifts. Here $r_{\rm drag}$ denotes the sound
horizon at the baryon drag epoch and is treated as a trainable parameter
of the reconstruction.

\item \textit{Cosmic chronometers (CC):} We additionally consider 31
model-independent measurements of the Hubble parameter $H(z)$ obtained
through the cosmic-chronometer method \cite{Jimenez:2001gg,Moresco:2024wmr},
covering the interval $0.09\leq z\leq1.965$.

\end{itemize}

Accordingly, we perform two independent reconstructions, using the
combined datasets PP+BAO+CC and PPS+BAO+CC. This allows us, in particular,
to examine the effect of the Cepheid calibration on the reconstructed
Barrow exponent and on the associated cosmological evolution.

\subsection{Network architecture and loss function}

We normalize the redshift according to
\begin{equation}
x=\frac{z}{z_{\max}},
\end{equation}
such that $x\in[0,1]$, and use it as the input variable of the network.
The primary reconstructed quantity is the Barrow entropy exponent
$\Delta(z)$. Following the Cosmo-PINN architecture
\cite{Paliathanasis:2026dqk},
and in order to obtain a stable reconstruction while suppressing
overfitting, we represent $\Delta$ through a finite Chebyshev expansion, namely
\begin{equation}
\label{eq:Delta_network}
\Delta(x)=A\tanh[g(x)],
\qquad
g(x)=\sum_{n=0}^{N}c_n T_n(2x-1),
\end{equation}
where $T_n$ are the Chebyshev polynomials and $c_n$ are trainable
coefficients. The hyperbolic-tangent map guarantees the bound
$|\Delta|<A$. In the present analysis we set $A=2$, thereby excluding
the singular limit $\Delta\rightarrow2$ directly through the network
architecture, without introducing an additional penalty term. The
Chebyshev coefficients are initialized at zero, and therefore the
training starts from $\Delta=0$, corresponding to the standard
Bekenstein-Hawking holographic dark-energy limit.

The dark-energy density parameter is reconstructed through the bounded
representation
\begin{equation}
\label{eq:Omega_network}
\Omega_{DE}(z)
=
\sigma\left[
{\rm logit}(1-\Omega_{m0})
+xN_{DE}(x)
\right],
\end{equation}
where $\sigma(y)=1/(1+e^{-y})$ is the logistic function and
$N_{DE}(x)$ denotes the corresponding neural-network output. This
construction automatically restricts $\Omega_{DE}$ to the physical
interval $(0,1)$ and, since the second term vanishes at $x=0$, imposes
$\Omega_{DE0}=1-\Omega_{m0}$. The Hubble function is subsequently
obtained from Eq.~(\ref{eq:Hubble_redshift}). Thus, the network reconstructs
the
relevant cosmological functions while respecting the physical bounds
already at the level of its architecture.

The training is controlled by the total loss function
\begin{equation}
\label{eq:loss_total}
\mathcal{L}_{\rm total}
=
\mathcal{L}_{\rm PDE}
+\mathcal{L}_{\rm DATA}
+\mathcal{L}_{\rm Priors}
+\mathcal{L}_{\rm SP},
\end{equation}
where the four contributions respectively encode the cosmological dynamics,
the observational information, soft priors on the cosmological
parameters, and smoothness conditions on the reconstructed functions.

The physics-informed contribution is defined as
\begin{equation}
\label{eq:loss_pde}
\mathcal{L}_{\rm PDE}
=
\frac{\lambda_{\rm PDE}}{N_c}
\sum_{i=1}^{N_c}
\mathcal{R}_{\rm PDE}^{2}(x_i),
\end{equation}
where $N_c$ denotes the number of collocation points and
$\mathcal{R}_{\rm PDE}(x_i)$ is the residual of the generalized Barrow
holographic dark-energy evolution equation (\ref{eq:OmegaDE_evolution}),
evaluated at
each collocation point $x_i$. Minimization of this contribution therefore
forces the reconstructed $\Delta(z)$ and $\Omega_{DE}(z)$ to satisfy the
underlying cosmological dynamics throughout the redshift interval, and
not merely at the locations of the observational measurements.

The observational contribution takes the form
\begin{equation}
\label{eq:loss_data}
\mathcal{L}_{\rm DATA}
=
\lambda_{\rm CC}\mathcal{L}_{\rm CC}
+\lambda_{\rm BAO}\mathcal{L}_{\rm BAO}
+\lambda_{\rm SN}\mathcal{L}_{\rm SN}.
\end{equation}
During training, the weights $\lambda_{\rm CC}$, $\lambda_{\rm BAO}$
and $\lambda_{\rm SN}$ are dynamically adjusted using the GradNorm
method, in order to balance the gradient contributions of the different
observational probes. Since the datasets contain substantially different
numbers of measurements, we normalize their contributions by the
corresponding number of data points. In particular, for the cosmic
chronometers we use
\begin{equation}
\label{eq:loss_cc}
\mathcal{L}_{\rm CC}
=
\frac{\chi_{\rm CC}^{2}}{N_{\rm CC}},
\end{equation}
whereas for BAO and SNIa we adopt the logarithmically compressed loss
\begin{equation}
\label{eq:loss_cap}
\mathcal{L}_{\rm Data}
=
\gamma^{2}
\ln\left[
1+
\frac{1}{\gamma^{2}}
\frac{\chi_{\rm Data}^{2}}{N_{\rm Data}}
\right],
\qquad
{\rm Data}\in\{{\rm BAO,SNIa}\}.
\end{equation}
The logarithmic form prevents very large initial values of
$\chi_{\rm Data}^{2}/N_{\rm Data}$ from dominating the optimization
during the early stages of training. In the analysis that follows we
set $\gamma=4$.

In addition to the reconstructed functions, the quantities
\begin{equation}
\theta\in\{H_0,\Omega_{m0},r_{\rm drag},C\}
\end{equation}
are treated as trainable cosmological parameters. Since their values are
coupled to the reconstruction of $\Delta(z)$ and $\Omega_{DE}(z)$, we
introduce soft Gaussian anchors through
\begin{equation}
\label{eq:loss_priors}
\mathcal{L}_{\rm Priors}
=
\sum_{\theta}
\frac{\lambda_{\theta}}{2}
\left(
\frac{\theta^{\rm PINN}-\bar{\theta}}
{\sigma_{\theta}}
\right)^2.
\end{equation}
Furthermore, for the training runs we adopt
\begin{equation}
(\bar{\Omega}_{m0},\sigma_{\Omega_m},\lambda_{\Omega_m})
=
(0.30,0.03,10^{-2}),
\end{equation}
\begin{equation}
(\bar H_0,\sigma_{H_0},\lambda_{H_0})
=
(68,2,10^{-4}),
\end{equation}
\begin{equation}
(\bar r_{\rm drag},\sigma_{r_{\rm drag}},\lambda_{r_{\rm drag}})
=
(147.1,3,10^{-4}),
\end{equation}
and
\begin{equation}
(\bar C,\sigma_C,\lambda_C)
=
(0.7,0.3,10^{-2}).
\end{equation}
These terms act as soft anchors rather than fixed constraints, allowing
the corresponding parameters to be adjusted by the combined
observational and physics-informed loss.

Finally, $\mathcal{L}_{\rm SP}$ contains weak smoothness penalties on
the first and second derivatives of $\Delta(z)$ and $\Omega_{DE}(z)$.
Their role is to suppress spurious high-frequency oscillations between
the collocation points without prescribing the overall redshift
dependence of the reconstructed functions. Hence, the functional form
of $\Delta(z)$ is not fixed a priori: its evolution emerges from the
combined action of the observational likelihood, the generalized Barrow
holographic dynamics, and the regularity conditions encoded in the
network.

\subsection{Posterior inference}

In order to quantify the uncertainties of the reconstructed cosmological
parameters and functions, we employ Hamiltonian Monte Carlo (HMC)
\cite{Neal:2011mrf}, using the No-U-Turn Sampler (NUTS)
\cite{JMLR:v15:hoffman14a}. The
posterior sampling is initialized from the optimized solution obtained
during the Cosmo-PINN training, such that the inference stage explores
the parameter space around a solution that already provides a
simultaneous description of the observational data and the underlying
cosmological dynamics.

The sampled quantities include the cosmological parameters
$\{\Omega_{m0},H_0,r_{\rm drag},C\}$, the Chebyshev coefficients
$\{c_n\}$ entering the reconstruction of $\Delta(z)$, and the weights
connecting the final hidden layer to the output layer of the network.
For the cosmological parameters we adopt Gaussian priors centered on
their corresponding values obtained from the trained Cosmo-PINN
solution, while the Chebyshev coefficients are assigned Gaussian priors
centered at zero.

For every posterior sample, the corresponding functions are propagated
through the trained network, thereby generating posterior distributions
not only for the cosmological parameters but also for the reconstructed
quantities, and in particular for $\Delta(z)$ and $\Omega_{DE}(z)$.
The same procedure is subsequently used to propagate the uncertainties
to derived cosmological quantities, such as $H(z)$, $w_{DE}(z)$ and the
deceleration parameter $q(z)$. We finally determine the $68\%$ and
$95\%$ credible intervals of the reconstructed functions from the
resulting posterior distributions.

 \begin{figure*}[!]
\centering\includegraphics[width=0.5\textwidth]{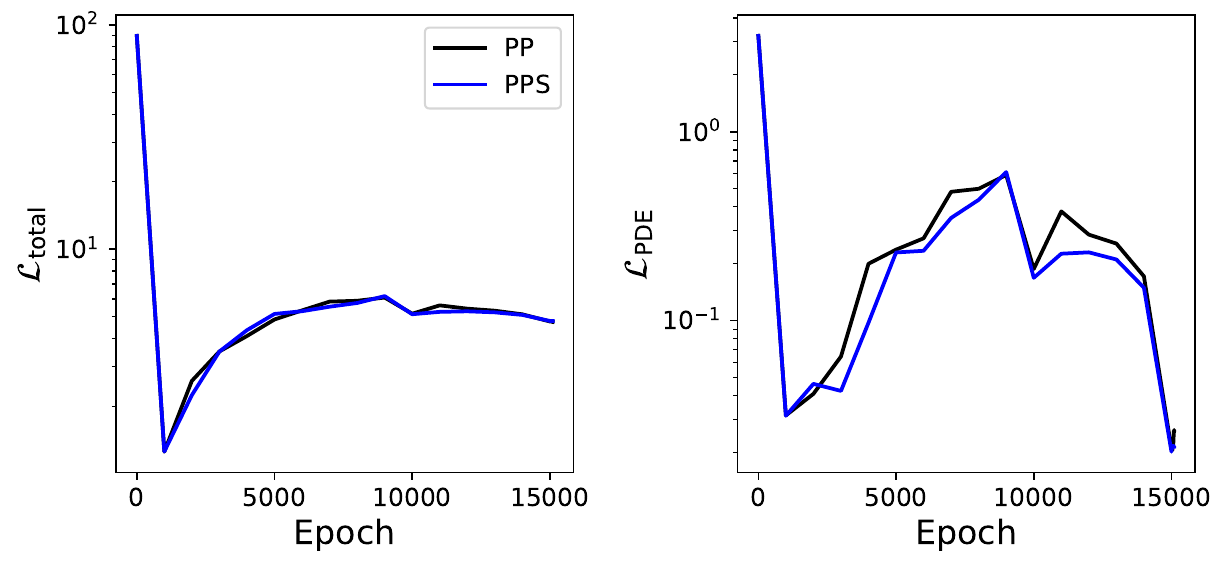}\\
 \includegraphics[width=0.8\textwidth]{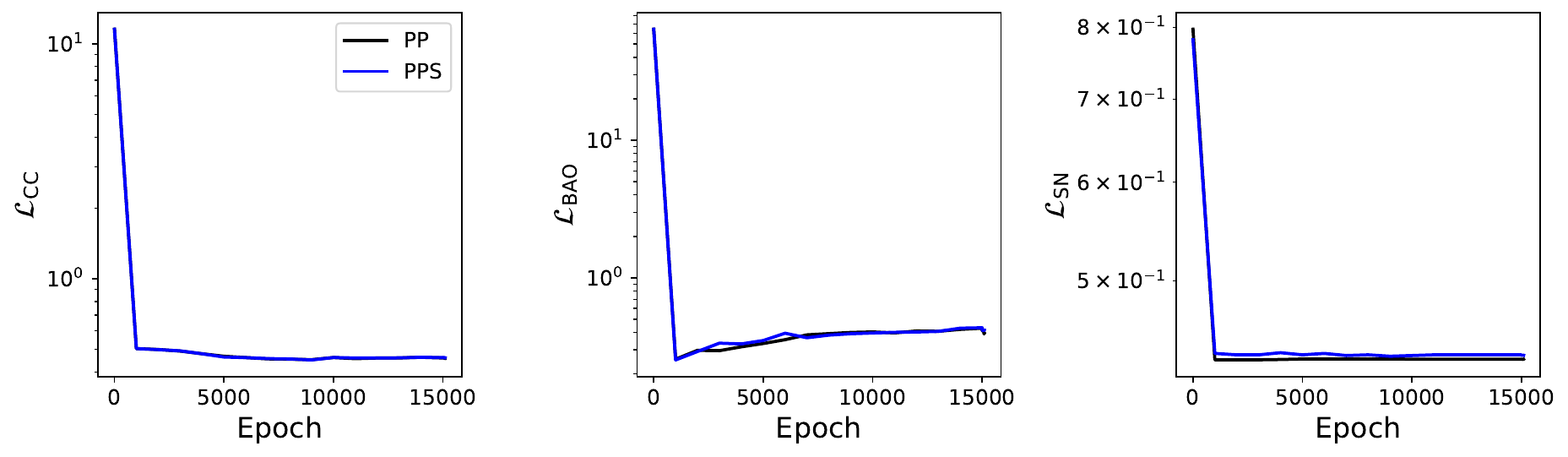}
\caption{\textit{
Evolution of the different loss components during training
for the PP+BAO+CC (black curves) and PPS+BAO+CC
(blue curves) dataset combinations.  Upper panels: total loss $\mathcal{L}_{\rm
total}$ and
physics-informed loss $\mathcal{L}_{\rm PDE}$. Lower panels: individual
observational loss contributions $\mathcal{L}_{\rm CC}$,
$\mathcal{L}_{\rm BAO}$, and $\mathcal{L}_{\rm SN}$. The first training phase
uses the Adam optimizer, while the second phase employs the L-BFGS optimizer.}}
\label{fig1}
\end{figure*}

\begin{figure*}[!]
\centering\includegraphics[width=0.7\textwidth]{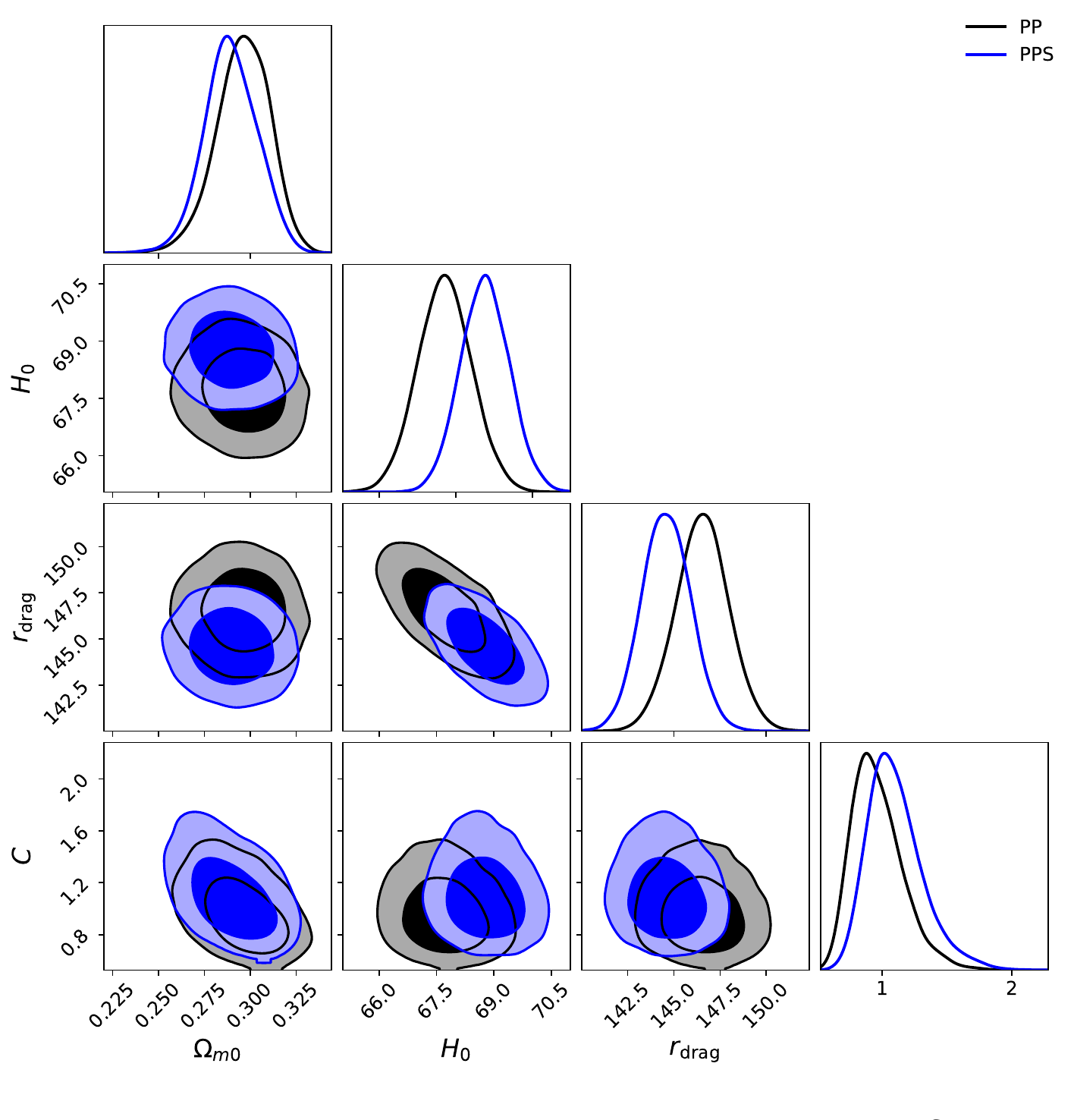}
\caption{\textit{One- and two-dimensional marginalized posterior distributions
for the cosmological parameters $H_0$, $\Omega_{m0}$, $r_{\rm drag}$, and
$C$, obtained from the HMC/NUTS analysis for the PP+BAO+CC and PPS+BAO+CC
dataset combinations. The contours correspond to the $68\%$ and $95\%$
credible regions.}}
\label{posteriorH0}
\end{figure*}

\begin{figure*}[!]
\centering\includegraphics[width=0.9\textwidth]{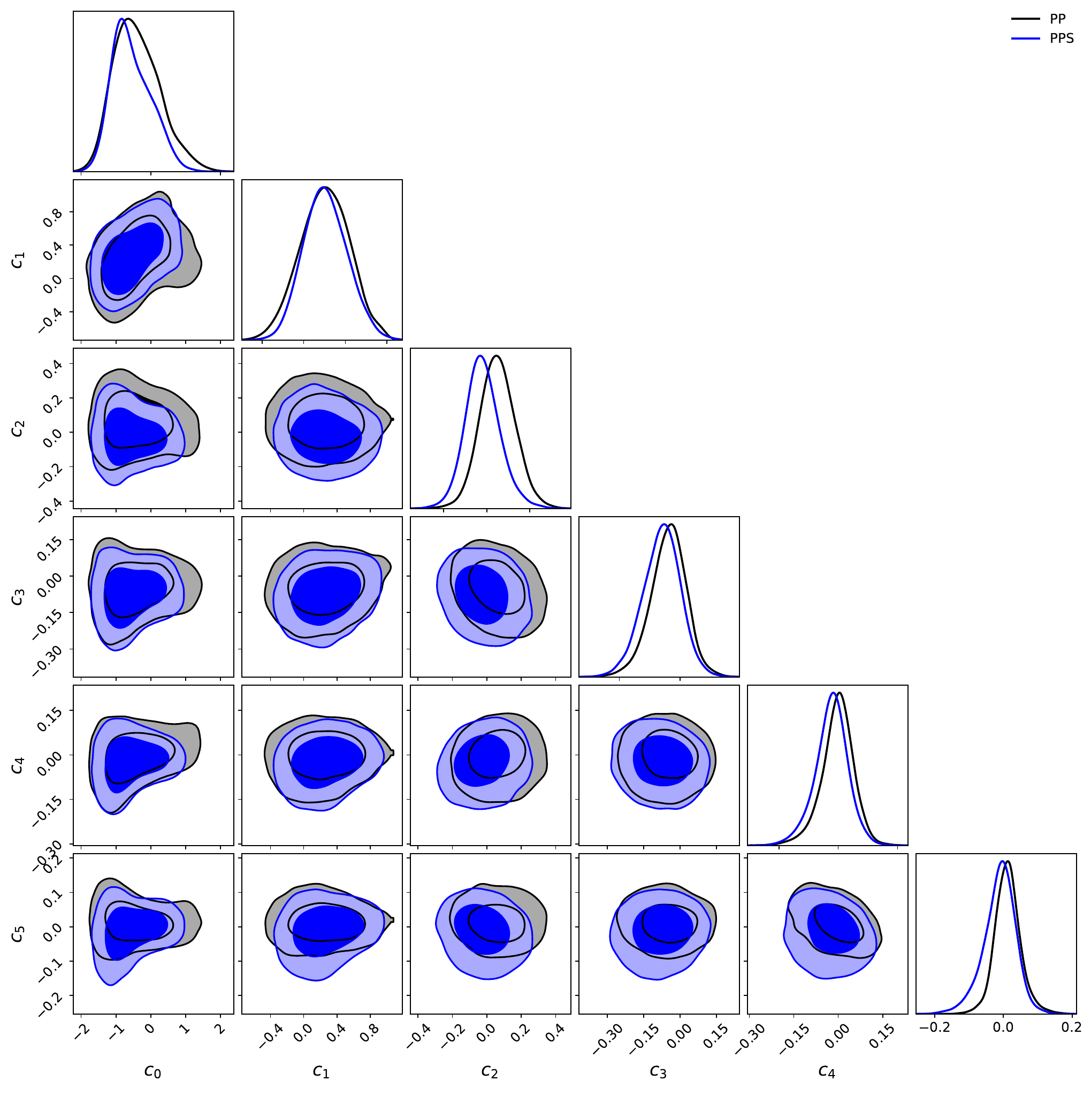}
\caption{\textit{One- and two-dimensional marginalized posterior distributions
for the Chebyshev coefficients $c_n$ entering the reconstruction of the
Barrow exponent $\Delta(z)$, for the PP+BAO+CC and PPS+BAO+CC dataset
combinations. The contours correspond to the $68\%$ and $95\%$ credible
regions.}}
\label{Chebyshev}
\end{figure*}

\begin{figure*}[!]
\centering\includegraphics[width=0.7\textwidth]{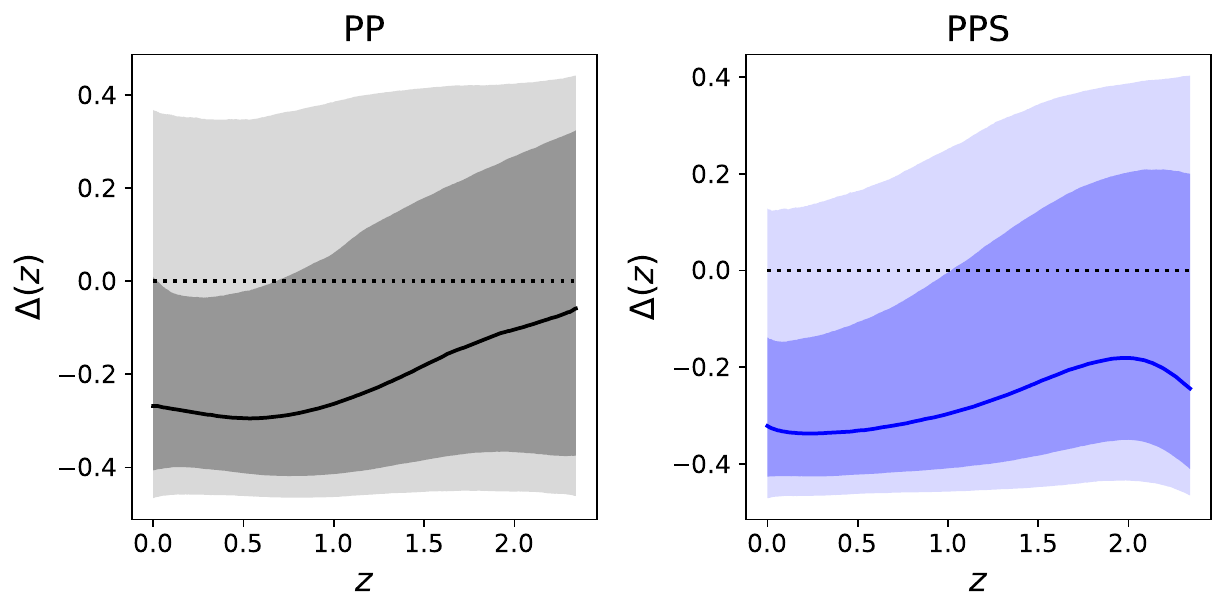}
\caption{\textit{Reconstructed generalized Barrow exponent $\Delta(z)$ for the
PP+BAO+CC (left panel) and PPS+BAO+CC (right panel) dataset combinations.
The solid curves denote the posterior mean reconstruction, while the shaded
regions correspond to the $68\%$ and $95\%$ credible intervals obtained
from the HMC/NUTS posterior sampling. The horizontal line at $\Delta=0$
indicates the standard holographic dark-energy limit.}}
\label{ReconstructedDelta}
\end{figure*}

\section{Reconstruction of the varying Barrow exponent and cosmological
implications}
\label{sec4}

Having introduced the theoretical framework and the corresponding
Cosmo-PINN implementation, we now proceed to the reconstruction of the
varying Barrow exponent and to the investigation of its cosmological
consequences. We first examine the training behavior of the network and the
posterior constraints on the cosmological parameters and on the coefficients
that determine the reconstructed function. We then focus on the central
result of the analysis, namely the data-driven reconstruction of
$\Delta(z)$, comparing the cases with and without the Cepheid calibration.
Finally, using the reconstructed Barrow exponent, we derive the corresponding
dark-energy equation-of-state parameter, the deceleration parameter, and the
evolution of the matter and dark-energy density parameters.

\subsection{Training and posterior constraints}

For the numerical reconstruction we employ a network with four hidden
layers, each containing 64 neurons, and use the hyperbolic tangent as the
activation function. The optimization is performed in two consecutive
stages. In the first stage we use the Adam optimizer for 15000 epochs,
allowing the network to explore the parameter space and approach the region
of the minimum. This is followed by 200 iterations of the L-BFGS optimizer,
which provides a more accurate local refinement of the solution. Moreover,
during the first 5000 Adam epochs we apply the GradNorm procedure in order
to dynamically adjust the relative weights
$\lambda_{\rm PDE}$, $\lambda_{\rm CC}$, $\lambda_{\rm BAO}$ and
$\lambda_{\rm SN}$ and balance the contributions of the physics and
observational terms to the optimization.

The evolution of the different loss components is presented in
Fig.~\ref{fig1}. In the upper panels we show the total loss
$\mathcal{L}_{\rm total}$ and the physics-informed contribution
$\mathcal{L}_{\rm PDE}$, while the lower panels display separately the
contributions associated with the CC, BAO and SNIa datasets. The behavior is
very similar for the PP+BAO+CC and PPS+BAO+CC combinations. During the early
part of the Adam optimization, all observational loss components decrease
rapidly from their initialization values and subsequently settle into a
slowly varying regime. In particular, the CC and SNIa contributions become
almost stationary relatively early in the training, whereas the BAO term
shows a somewhat longer adjustment before reaching its final plateau.

The evolution of $\mathcal{L}_{\rm PDE}$ is slightly different, reflecting
the competition between fitting the observations and enforcing the
generalized Barrow holographic evolution equation. After its initial rapid
decrease, the physics residual increases during part of the Adam stage as
the network readjusts its parameters in order to accommodate simultaneously
the different data contributions and the physical constraint. Consequently,
$\mathcal{L}_{\rm total}$ also increases after its early minimum and then
approaches a stable regime. Once the L-BFGS refinement is activated, the
physics residual decreases sharply, while the observational contributions
remain essentially unchanged. This behavior shows that the second
optimization stage improves the satisfaction of the dynamical equation
without spoiling the description of the data. Hence, the final trained
configuration represents a compromise in which both observational and
physics-informed requirements are simultaneously fulfilled.

We next examine the posterior distributions obtained through the HMC/NUTS
analysis described in Sec.~\ref{sec3} above. In Fig.~\ref{posteriorH0} we
present the
one- and two-dimensional marginalized posterior distributions for
$\Omega_{m0}$, $H_0$, $r_{\rm drag}$ and $C$, for the two observational
combinations. The matter density parameter is very similar in the two
reconstructions, with strongly overlapping posterior regions. The same is
true for the holographic parameter $C$, whose posterior distributions are
broad and largely compatible between PP and PPS. On the other hand, the
inclusion of the Cepheid calibration produces the expected shift in the
distance-scale parameters: the PPS reconstruction favors a larger value of
$H_0$ and, correspondingly, a smaller value of $r_{\rm drag}$ than the PP
case. The joint distributions also exhibit the characteristic
anti-correlation between $H_0$ and $r_{\rm drag}$, reflecting the way in
which the absolute distance calibration is accommodated by the BAO and
supernova measurements.

Additional information on the functional reconstruction is provided in
Fig.~\ref{Chebyshev}, where we show the posterior distributions of the Chebyshev
coefficients $c_n$ entering  (\ref{eq:Delta_network}). The coefficients
of the lowest-order modes possess the broadest posterior distributions and
therefore carry most of the freedom required to describe the reconstructed
redshift dependence. In contrast, as the polynomial order increases, the
corresponding coefficients become progressively more concentrated around
zero. In particular, the higher-order modes are strongly suppressed for
both observational combinations. This behavior is important, since it shows
that the reconstruction is dominated by smooth, low-order variations rather
than by high-frequency oscillatory structure. It therefore provides an
additional indication that the redshift dependence obtained below is not
driven by poorly constrained high-order modes or by an overfitting of the
observational data.

Having established the stability of the training procedure and the
posterior behavior of both the cosmological and reconstruction parameters,
we can now turn to the central result of the analysis, namely the inferred
redshift evolution of the Barrow exponent itself.

\subsection{Reconstruction of the Barrow exponent}
\label{sec4B}

We now turn to the central result of the analysis, namely the reconstruction
of the redshift dependence of the Barrow exponent. In
Fig.~\ref{ReconstructedDelta} we present
the
reconstructed $\Delta(z)$ together with its $68\%$ and $95\%$ credible
intervals for the PP+BAO+CC and PPS+BAO+CC combinations. In both cases the
posterior mean is found in the negative-$\Delta$ region over the whole
redshift interval considered, although the statistical uncertainties are
sufficiently large that the standard holographic limit $\Delta=0$ remains
compatible with the reconstruction within the broader credible region.

For the PP combination, the reconstructed mean exhibits a mild evolution:
starting from a negative value at the present epoch, it decreases slightly
towards intermediate redshifts and subsequently evolves towards values
closer to zero as the redshift increases. Nevertheless, the corresponding
credible intervals remain relatively broad and therefore do not provide
evidence for a statistically significant departure from a constant Barrow
exponent. In particular, both the standard holographic value $\Delta=0$ and
constant negative values of $\Delta$ remain compatible with the
reconstruction over substantial parts of the redshift interval.

The inclusion of the Cepheid calibration has a visible effect on the
reconstructed exponent. As shown in the right panel of
Fig.~\ref{ReconstructedDelta}, the PPS+BAO+CC combination shifts the posterior
mean of
$\Delta(z)$ further towards negative values, especially at low and
intermediate redshifts. The reconstructed function again displays a smooth
redshift evolution, increasing towards less negative values at higher
redshift before turning slightly downwards near the upper edge of the
reconstruction range. However, the uncertainties increase considerably
towards high redshift, and thus the detailed behavior of the mean curve in
this region should not be interpreted as evidence for a specific evolution.

It is worth emphasizing that these results have been obtained without
assuming any of the linear, CPL-like, exponential or other parametrizations
previously considered for a varying Barrow exponent. Instead, the shape of
$\Delta(z)$ emerges from the combined observational information and the
generalized Barrow holographic evolution equation. The suppression of the
higher-order Chebyshev coefficients discussed in the previous subsection
further indicates that the reconstructed behavior is mainly controlled by
the lowest-order modes and is therefore smooth rather than being generated
by rapid oscillations of the functional basis.

Hence, the reconstruction reveals a mild preference of the posterior mean
for negative values of the generalized Barrow exponent, with this tendency
becoming stronger when the Cepheid calibration is included. At the same
time, the present late-time data do not require a departure from the
standard holographic limit, nor do they establish a statistically
significant redshift variation of $\Delta$. Rather, they delimit the range
of functional behaviors of the Barrow exponent that are compatible
simultaneously with the observations and with the holographic cosmological
dynamics.

\begin{figure*}[!]
\centering\includegraphics[width=0.7\textwidth]{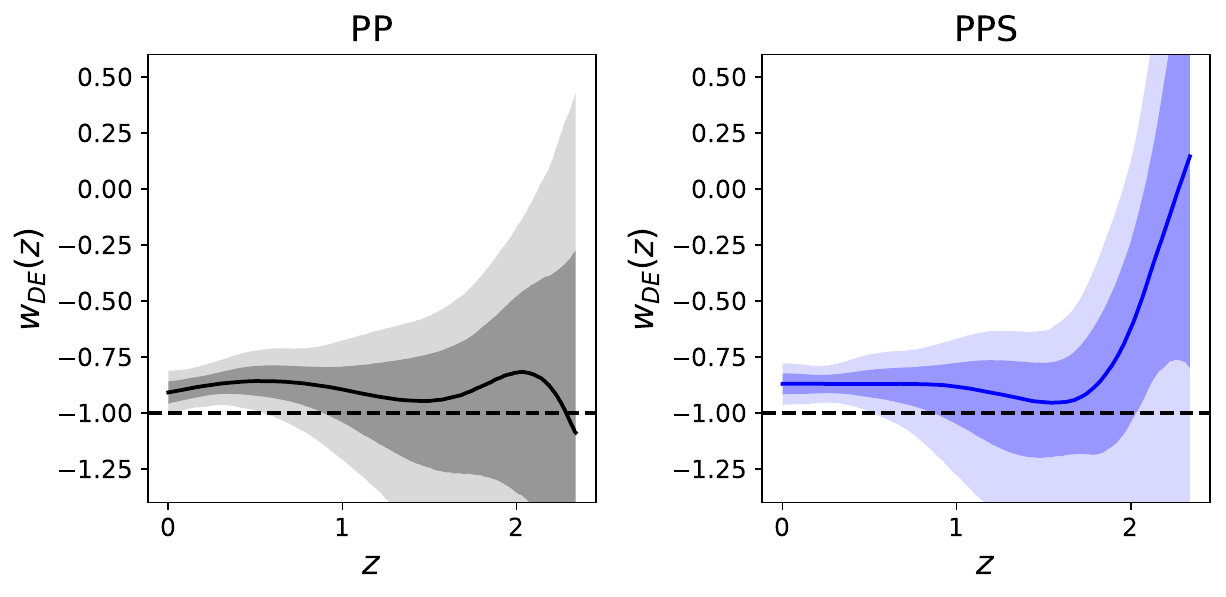}
\caption{\textit{Reconstructed dark-energy equation-of-state parameter
$w_{DE}(z)$ for the PP+BAO+CC (left panel) and PPS+BAO+CC (right panel)
dataset combinations. The solid curves denote the posterior mean
reconstruction, while the shaded regions correspond to the $68\%$ and
$95\%$ credible intervals. The horizontal line at $w_{DE}=-1$ indicates
the cosmological-constant value and allows for a direct assessment of
possible quintessence-like, phantom-like, and phantom-divide-crossing
behavior.}}
\label{Reconstructedwde}
\end{figure*}

\subsection{Cosmological implications}
\label{sec4C}

Having reconstructed the Barrow exponent, we can investigate the
corresponding cosmological evolution without introducing any additional
parametrization. We first consider the effective dark-energy
equation-of-state parameter, which follows from Eq.~(\ref{eq:wDE_redshift}) and
is
shown in Fig.~\ref{Reconstructedwde}. At low redshifts the central
reconstructions for
both observational combinations lie close to the cosmological-constant
value $w_{DE}=-1$. For the PP combination, $w_{DE}(z)$ exhibits only a mild
evolution throughout most of the reconstructed range, while the uncertainty
band progressively broadens towards higher redshift. Consequently, both
quintessence-like behavior, $w_{DE}>-1$, and phantom behavior,
$w_{DE}<-1$, are allowed within the posterior region, and a crossing of the
phantom divide is not excluded.

\begin{figure*}[!]
\centering\includegraphics[width=0.7\textwidth]{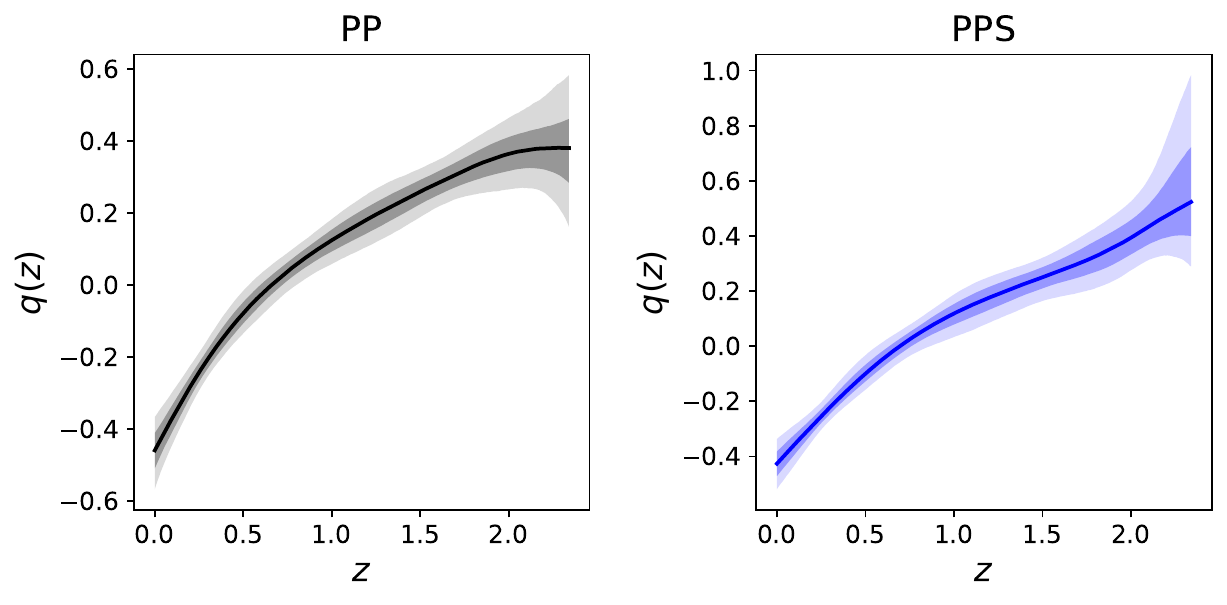}
\caption{\textit{Reconstructed deceleration parameter $q(z)$ for the
PP+BAO+CC (left panel) and PPS+BAO+CC (right panel) dataset combinations.
The solid curves denote the posterior mean reconstruction, while the shaded
regions correspond to the $68\%$ and $95\%$ credible intervals obtained
from the HMC/NUTS posterior sampling. The horizontal line at $q=0$ marks
the transition between decelerated and accelerated expansion.}}
\label{Reconstructedq}
\end{figure*}

\begin{figure*}[!]
\centering\includegraphics[width=0.7\textwidth]{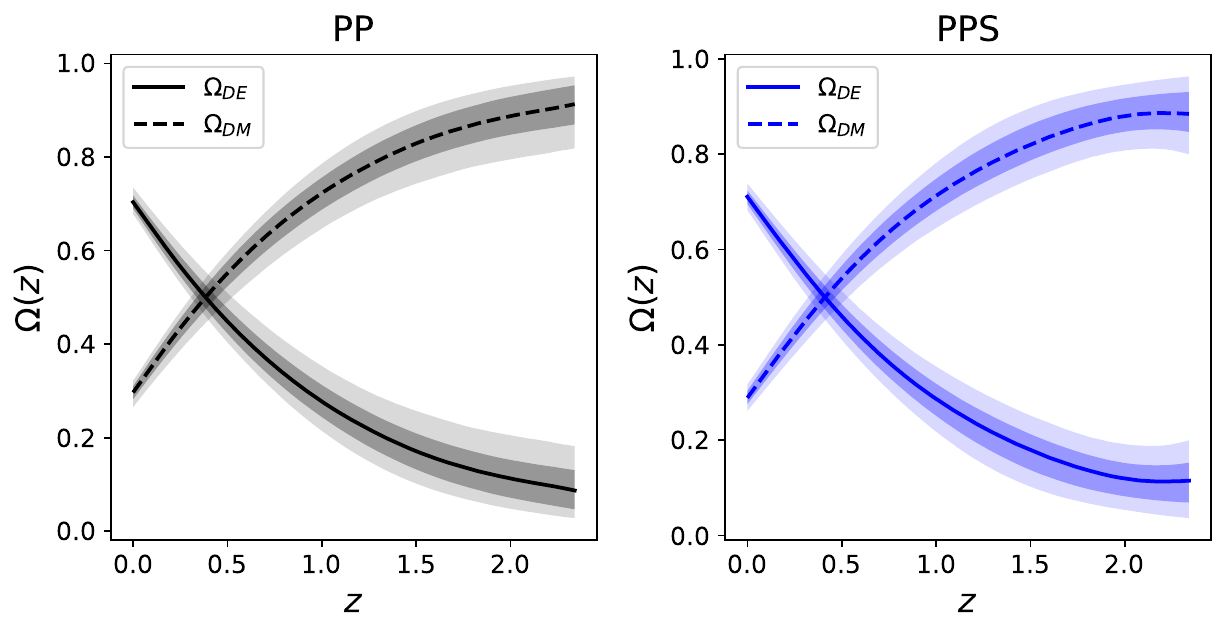}
\caption{\textit{Reconstructed evolution of the dark-energy and matter
density parameters, $\Omega_{DE}(z)$ and $\Omega_m(z)$, respectively, for
the PP+BAO+CC (left panel) and PPS+BAO+CC (right panel) dataset
combinations. The solid curves denote the posterior mean reconstructions,
while the shaded regions correspond to the $68\%$ and $95\%$ credible
intervals obtained from the HMC/NUTS posterior sampling. The crossing of
$\Omega_{DE}(z)$ and $\Omega_m(z)$ marks the transition from matter to
dark-energy domination.}}
\label{ReconstructedOm}
\end{figure*}

A similar picture is obtained for the PPS combination at low and
intermediate redshifts. The posterior mean remains close to $-1$ up to
$z\sim {\cal O}(1)$, before developing a stronger evolution at larger
redshift. Nevertheless, the credible intervals become rapidly wider in
this regime, and therefore the pronounced high-redshift evolution of the
central curve should be interpreted with caution. The robust conclusion is
that the reconstructed generalized Barrow holographic scenario can
accommodate an effective dark-energy sector close to the cosmological
constant at late times, while allowing both quintessence-like and
phantom-like realizations within the reconstructed uncertainties.

The corresponding deceleration parameter is presented in
Fig.~\ref{Reconstructedq}. For both PP and PPS, its present value is negative,
confirming the accelerated expansion of the Universe. Moving towards higher
redshift, $q(z)$ increases monotonically and crosses zero at a redshift below
$z=1$, signaling the transition from the recent
accelerated phase to the preceding decelerated expansion. The two
observational combinations lead to very similar transition histories,
although the PPS reconstruction presents slightly larger uncertainties at
the upper end of the redshift interval. Thus, despite the freedom allowed
in the redshift dependence of $\Delta(z)$, the reconstructed model naturally
recovers the standard sequence of a past decelerating epoch followed by the
current accelerated phase.

Finally, in Fig.~\ref{ReconstructedOm} we display the reconstructed evolution
of
the
dark-energy and matter density parameters, $\Omega_{DE}(z)$ and
$\Omega_m(z)$, respectively. In both observational combinations the
present Universe is dark-energy dominated, with
$\Omega_{DE0}\simeq 0.7$ and $\Omega_{m0}\simeq0.3$, while towards higher
redshift the matter fraction grows and the dark-energy contribution
decreases. The two components become comparable at an intermediate
redshift, after which matter dominates the cosmic energy budget. The PP
and PPS reconstructions lead to very similar qualitative histories,
although small quantitative differences are induced by the different
absolute-distance calibration.

The combined behavior displayed in
Figs.~\ref{Reconstructedwde}-\ref{ReconstructedOm} therefore
shows that the data-driven reconstruction of the Barrow exponent produces
a cosmologically viable late-time history. In particular, it yields a
dark-energy sector close to $w_{DE}=-1$ at low redshift, a transition from
deceleration to acceleration at recent times, and the expected succession
from matter domination to dark-energy domination. These properties are
obtained without prescribing the redshift dependence of $\Delta$ in
advance, but instead emerge from the physics-informed reconstruction of the
underlying entropy deformation.

\section{Conclusions}
\label{Conclusions}

The nature of dark energy and its possible connection with fundamental
properties of gravity and horizon thermodynamics remain among the central
open questions in cosmology. Holographic dark energy provides an interesting
framework in this direction, relating the dark-energy density to the
infrared structure of the theory, while generalized entropy relations allow
possible quantum-gravitational modifications of horizon thermodynamics to
be incorporated at the cosmological level. In particular, in Barrow
holographic dark energy one uses the exponent $\Delta$, which quantifies
the deformation from the standard Bekenstein-Hawking entropy. If such a
deformation originates from quantum-gravitational effects, however, there
is no fundamental reason for its magnitude to remain constant throughout
cosmic evolution, motivating the generalized scenario in which $\Delta$
becomes redshift dependent.

Previous investigations of this possibility have relied on specific
parametrizations for $\Delta(z)$, which inevitably introduce some dependence
on the assumed functional form. In the present work we followed the opposite
strategy and reconstructed its evolution directly from cosmological
observations. For this purpose we employed the Cosmo-PINN framework,
incorporating the generalized Barrow holographic evolution equation directly
into the neural-network training. Hence, the reconstructed function is
simultaneously constrained by observations and required to satisfy the
underlying cosmological dynamics. We considered PantheonPlus supernovae,
with and without the Cepheid calibration, together with DESI DR2 BAO and
cosmic-chronometer measurements, and used HMC/NUTS sampling to determine the
posterior uncertainties of the cosmological parameters and reconstructed
functions.

The training and posterior analysis showed that the physics-informed
reconstruction is stable for both observational combinations. The
observational loss components converge during the Adam optimization, while
the subsequent L-BFGS stage significantly improves the satisfaction of the
generalized Barrow holographic evolution equation without degrading the
description of the data. Furthermore, the posterior distributions of the
Chebyshev coefficients exhibit a progressive suppression of the higher-order
modes, indicating that the inferred redshift dependence is predominantly
controlled by smooth, low-order variations rather than by high-frequency
oscillations. The two dataset combinations lead to compatible values of
$\Omega_{m0}$ and the holographic parameter $C$, while inclusion of the
Cepheid calibration produces the expected increase of $H_0$ and corresponding
decrease of $r_{\rm drag}$.

The central result of the analysis is the reconstruction of the generalized
Barrow exponent itself. For both PP+BAO+CC and PPS+BAO+CC, the posterior mean
of $\Delta(z)$ lies in the negative region over the reconstructed redshift
interval and displays a mild and smooth evolution. This tendency becomes
more pronounced when the Cepheid calibration is included, particularly at
low and intermediate redshifts. Nevertheless, the current uncertainties do
not allow us to claim evidence either for a departure from the standard
holographic limit $\Delta=0$ or for a genuinely evolving Barrow exponent.
Constant negative values of $\Delta$ also remain compatible with the
reconstruction. Thus, the important result is not a detection of an evolving
entropy deformation, but the direct determination of the functional
behavior of $\Delta(z)$ allowed simultaneously by observations and by the
generalized Barrow holographic dynamics, without prescribing its redshift
dependence beforehand.

The reconstructed exponent leads to a consistent and phenomenologically
viable late-time cosmological evolution. The effective dark-energy
equation-of-state parameter remains close to the cosmological-constant value
at low redshift, while both quintessence-like and phantom-like behavior, as
well as phantom-divide crossing, are allowed within the reconstructed
credible regions. At higher redshift the uncertainties increase and the
detailed evolution should therefore be interpreted cautiously. Moreover,
the deceleration parameter exhibits the expected transition from a past
decelerating phase to the current accelerated epoch, while the density
parameters reproduce the standard evolution from matter domination to
dark-energy domination. Hence, allowing the entropy deformation itself to
be inferred from observations preserves the successful background
cosmological evolution while considerably relaxing assumptions about its
underlying functional form.

Moreover, the novelty of the present analysis lies also
in the nature of the reconstructed quantity. Cosmological reconstruction
techniques are commonly applied to observables or effective quantities such
as $H(z)$ or $w_{DE}(z)$. Here, instead, the data are used to reconstruct a
quantity entering the underlying entropy law. The analysis therefore
illustrates how physics-informed machine learning can connect cosmological
observations with quantities characterizing the more fundamental theoretical
structure of a model. In this sense, late-time observations can directly
probe the possible evolution of the Barrow entropy deformation, while the
fact that $\Delta=0$ remains allowed shows that present data constrain, but
do not require, a departure from the standard Bekenstein-Hawking
holographic framework.

Several extensions would be interesting. Future observations with increased
precision and redshift coverage could determine whether the mild evolution
suggested by the posterior mean is a genuine feature or reflects the present
statistical freedom. Incorporating cosmological perturbations and
structure-growth information would provide complementary constraints on the
reconstructed dynamics, while alternative infrared cutoffs and other
generalized entropy constructions could be investigated within the same
framework. More generally, the strategy developed here can be applied to
cosmological scenarios in which quantities usually treated as constants or
assigned phenomenological parametrizations may instead possess an underlying
scale or time dependence. Physics-informed reconstruction may therefore
allow cosmological observations not only to constrain the parameters of a
theory, but also to reveal aspects of its functional structure.
\\
\textbf{Data Availability Statement: }Data are available on reasonable request.
\\
\textbf{Code Availability Statement: }Code will be made available on reasonable
request.
\\
\textbf{Conflict of interest: }The authors declare that there is no conflict of
interest

\begin{acknowledgments}
This work was partially supported by FONDECYT Grant 1240514. The authors would
like to acknowledge the
contribution of the LISA CosWG, and of COST Actions CA21136 ``Addressing
observational tensions in cosmology with systematics and fundamental physics
(CosmoVerse)'', CA21106 ``COSMIC WISPers in the Dark Universe: Theory,
astrophysics and experiments (CosmicWISPers)'', and CA23130 ``Bridging high and
low energies in search of quantum gravity (BridgeQG)''.
\end{acknowledgments}

\bibliography{references}

\end{document}